\documentclass[prl,preprintnumbers,superscriptaddress,shownopacs,twocolumn,balancelastpage,floatfix]{revtex4}

\usepackage{amsmath,amssymb,amsfonts}
\usepackage{bm,graphicx,color}

\newcommand{\expval}[1]{\langle{#1}\rangle}

\newcommand{\jstat}{\bar{\jmath}}
\newcommand{\IV}{$I$-$V$}

\begin{document}

  \title{Dielectric breakdown of Mott insulators in 
  dynamical mean-field theory}

  \author{Martin Eckstein}
  \affiliation{Theoretical Physics, ETH Zurich, 8093 Zurich, Switzerland}  

  \author{Takashi Oka}
  \affiliation{Department of Physics, University of Tokyo, Hongo, Tokyo 113-0033,
  Japan}

  \author{Philipp Werner}
  \affiliation{Theoretical Physics, ETH Zurich, 8093 Zurich, Switzerland}

  \date{\today}

\begin{abstract}
Using nonequilibrium dynamical mean-field theory, we compute the time evolution of the 
current in a Mott insulator after a strong electric field is turned on. We observe the 
formation of a quasistationary state in which the current is almost time-independent 
although the system is constantly excited. At moderately strong fields this state is 
stable for quite long times. The stationary current exhibits a threshold behavior as
a function of the field, in which the threshold increases with the Coulomb interaction 
and vanishes as the metal-insulator transition is approached.
\end{abstract}

\maketitle


Nonequilibrium phase transitions and nonlinear transport
are becoming central issues in the study of  strongly correlated
systems. One of the most basic phenomena is the dielectric breakdown  
(destruction
of insulating states due to strong electric fields)  
\cite{tag,Oka2003a,Oka2010a}.
In Mott insulators, the electron motion is frozen as a result
of strong repulsive interactions \cite{Imada1998},
and in equilibrium the doping of carriers into a Mott insulator
leads to interesting quantum states such as
high-$T_c$ superconductivity in two-dimensional materials.
Thus, it is natural to ask how nonequilibrium carriers behave when
electrons in a Mott insulator start to move in response to strong  
electric fields.

Experimentally, the physics of nonlinear transport
in correlated electron systems  has been studied in
oxides \cite{tag} as well as in organic materials \cite{EXPorganic}.
In one dimensional Mott insulators,
dielectric breakdown was observed and it was found that the
current increases with a threshold behavior \cite{tag}.
The current-voltage (\IV) characteristics 
exhibits
a strong non-linearity with a negative differential resistivity between 
the weak current and  large current regimes.
More recently, the problem of nonlinear transport is attracting  
interest also in the cold atom community,
where a novel realization of the Mott insulating state has been  
achieved.
In Ref.~\cite{Greiner2002} the effect of a potential gradient was  
studied to
probe the excitation spectrum.

A theoretical description of nonlinear transport
is challenging because one needs to take
into account two nonperturbative effects, electric fields
and electron-electron interactions, simultaneously.
In one-dimensional systems, reliable numerical techniques such as
exact diagonalization and the time-dependent density matrix  
renormalization group 
are available,
and a threshold behavior in the
current was indeed observed \cite{Oka2003a}. A
many-body Schwinger-Landau-Zener mechanism, in which doubly occupied  
states (doublons)
and holes are pair produced by quantum tunneling
was proposed as an explanation \cite{Oka2010a}.
However, in previous studies,
a direct calculation of the \IV-characteristics
is absent since a steady state current cannot be easily  
reached in a finite system.
This made it difficult to compare theory with experiments.
Another totally unexplored issue is
the temperature effect,
where experiments suggest a relatively strong temperature
dependence of the current.

The purpose of the present paper is to address these questions using 
dynamical mean-field theory (DMFT)
\cite{Georges1996}, which is suitable for the study of high-dimensional
bulk systems. Nonequilibrium DMFT \cite{Schmidt2002} has been used to reveal
various types of steady states and relaxation phenomena in the Falicov-Kimball 
model \cite{Freericks2006a,Eckstein2008a,Tsuji2009a} and the Hubbard model
\cite{Eckstein2009a}. For the current analysis one must compute the dynamics
of the Hubbard model at rather strong interactions up to relatively long times.
So far, this task has been prohibitively difficult for impurity solvers based on 
real-time quantum Monte Carlo \cite{Eckstein2009a}, but it has become accessible 
recently through an implementation of the self-consistent hybridization expansion 
within the Keldysh framework \cite{Eckstein2010c}.


In the following we focus on the Mott insulating phase in the half-filled 
Hubbard model on a $d$-dimensional cubic lattice with lattice spacing $a$,
\begin{equation}
\label{hubbard}
H
=
\sum_{\langle ij \rangle\sigma} 
V_{ij}(t) \,
c_{i\sigma}^\dagger c_{j\sigma}
+
U\sum_i (n_{i\uparrow}-\tfrac12)(n_{i\downarrow}-\tfrac12).
\end{equation}
Here $c_{i\sigma}^\dagger$ ($c_{i\sigma}$) denotes the creation (annihilation) 
operator for an electron with spin $\sigma$ at lattice site ${\bm R}_i$, $U$ 
is the local Coulomb repulsion, and $V_{ij}$ describes hopping between the 
sites. To study the dielectric breakdown of the Mott insulator, we initially 
prepare the system 
in thermal equilibrium
at temperature $T=1/\beta$ and apply a spacially homogeneous
electric field ${\bm F}(t)$ for time $t>0$. Using a gauge with pure vector 
potential ${\bm A}(t)$, i.e., ${\bm F}(t)=-\partial_t {\bm A(t)}/c$, 
${\bm F}(t)$ is incorporated into Eq.~(\ref{hubbard}) by means of the Peierls 
substitution, $V_{ij}(t)=V_{ij}^0 \exp[ie ({\bm R}_j-{\bm R}_i){\bm A}(t)/\hbar c]$.
The field is chosen to point along the body diagonal $\hat{\bm\eta}=(1\ldots1)^t$
of the unit cell. It is turned on to a value $F$ within a switching time $t_0$, 
$\bm F(t) = \hat{\bm \eta} F\, r(t/t_0)$, using a switching profile $r(x)= \tfrac12-
\tfrac34\cos(\pi x)+\tfrac14 \cos(\pi x)^3$ for $0\le x\le 1$. The main results 
turn out to be independent of the switching, and we choose $t_0=3$ if 
not stated otherwise.

In the limit $d=\infty$ \cite{Metzner1989a} the problem is solved exactly using 
nonequilibrium DMFT. The DMFT self-consistency equations for 
finite electric field are identical for the Hubbard model and the Falicov-Kimball 
model, and they are detailed in Ref.~\cite{Freericks2006a}. The DMFT single-site problem, 
on the other hand, is identical to the zero field case, and we solve it by means of the 
self-consistent hybridization expansion 
\cite{Eckstein2010c} up to second order (one-crossing approximation), which 
is known to be reliable for the insulating phase and the crossover regime 
\cite{Pruschke1993a,Eckstein2010c}. We choose units of energy such 
that the density of states is given by $\rho(\epsilon) \propto \exp(-\epsilon^2)$, with 
a width (second moment) $W=\sqrt{2}/2$. Time and field are measured in units 
of $1/\sqrt{2}W$ and $\sqrt{2} W/ea$, respectively ($\hbar=1$). In equilibrium, 
a first-order Mott transition is found in the paramagnetic phase \cite{Georges1996},
with a critical end-point at $U_c\approx 3.1$ and $T_c \approx 0.02$ (determined 
within the one-crossing approximation).


\begin{figure}
\includegraphics[angle=0,width=0.99\columnwidth]{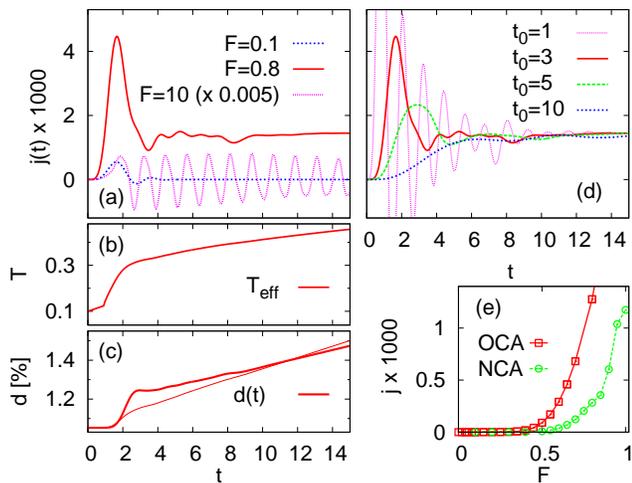}
 \caption{
 Time-evolution after turning on a field $F$ in the insulator ($U=5$, 
 $\beta=10$): (a) The current $j(t)$; results for $F=10$  are multiplied 
 with a factor $0.005$ to match the scale. (b) Effective temperature
 $T_\text{eff}$ and (c), double occupancy $d(t)$ for $F=0.8$. 
 The thin line shows the thermal  expectation value of $d$ at temperature 
 $T=T_\text{eff}(E(t))$. (d) Current at $F=0.8$ for various switch-on times 
 $t_0$. (e) Current, averaged for $t\ge 10$, obtained using either NCA or OCA to 
 solve the DMFT impurity problem.
 }
\label{fig01}
\end{figure}

After an electric field is turned on in the Mott-insulating phase, we either observe
the formation of a quasistationary state with time-independent current (Fig.~\ref{fig01}a),
or, for very large values of $F$, the emergence of $2\pi/F$-periodic Bloch oscillations ($F=10$ 
in Fig.~\ref{fig01}a). In analogy to the Falicov-Kimball model \cite{Freericks2006a}, where 
Bloch oscillations are quenched by the interaction, there is no sharp separation between 
oscillatory and non-oscillatory regimes, but instead the current behaves irregularly at 
intermediate fields. In the following, we only study the quasistationary state, which 
will reveal the dielectric breakdown of the Mott insulator at moderately large, 
possibly experimentally accessible fields.

Since our system is not coupled to a thermal bath, the energy $E=\expval{H(t)}$ increases 
at a rate $\dot{E} = F(t) j(t)$, and thus a stationary state with nonzero current cannot 
exist forever. However, we find that $j(t)$ remains remarkably stable even after a considerable 
energy increase. This fact becomes clearly evident if one looks at the effective 
temperature instead of the energy, i.e., the temperature $T_\text{eff}(E)$ after a hypothetical 
thermalization at energy $E$: For $F=0.8$ in Fig.~\ref{fig01}, e.g., $T_\text{eff}$ increases 
by a factor $1.5$ during times $4 < t < 15$ (Fig.~\ref{fig01}b), while $j(t)$ remains almost 
constant. The saturation of the current and a simultaneous increase in the double occupancy 
$d=\expval{n_{i\uparrow}n_{i\downarrow}}$ (Fig.~\ref{fig01}c) indicate that the current 
flow causes excitations in the system that are immobile, just like 
the spin fluctuations in the ground state that lead to a finite double occupancy but no 
linear response conductivity. These excitations do not thermalize on the timescale of our 
simulation. Otherwise the current would increase thermally, and $d(t)$ would have 
to match its expectation value in thermal equilibrium at temperature $T_\text{eff}(E(t))$ 
(thin line in Fig.~\ref{fig01}c), which is not the case. This behavior is consistent with recent 
experiments on ultracold gases \cite{Strohmeier2010a}, where it was found that artificially 
created double occupancies in the Mott insulator relax only on exponentially long timescales.

The quasistationary current turns out to be more or less independent on how the field 
is turned on, and by increasing the switching time $t_0$ one can reduce the otherwise rather 
strong transient oscillations (Fig.~\ref{fig01}d). Even for slow switching, however, the 
transient current can be orders of magnitude larger than the stationary current. This fact
is already entailed in the linear response relation $j(t)= \int_0^{t}\!ds \,\sigma(s) F(t-s)$ 
which always holds for small enough times. The transient is at least proportional 
to $F$, while the long-time limit can be exponentially 
small (see below). In particular, in the Mott insulator at $T=0$ the dc conductivity 
$\sigma_\text{dc}=\int_0^{\infty}\! ds\, \sigma(s)$ vanishes, while the integral 
$\alpha = \int_0^{\infty} \!dt \int_0^{t}\!ds\,\sigma(s)$ over the current yields a 
nonzero static polarizability.

After averaging over times $t\ge 10$, the quasistationary current $\jstat$ at $U=5$
shows a sharp increase around $F = 0.5$ (Fig.~\ref{fig01}e), which is the 
hallmark of the dielectric breakdown. We note that both first order (NCA) and second 
order (OCA) implementations of the self-consistent hybridization expansion yield 
similar results, but in analogy to equilibrium calculations the 
insulating behavior is overestimated by NCA, such that the increase of the current 
is shifted to stronger fields. In the following we stick to the more reliable 
OCA as an impurity solver.

\begin{figure}
\end{figure}
\begin{figure}
 \centerline{\includegraphics[angle=0,width=0.92\columnwidth]{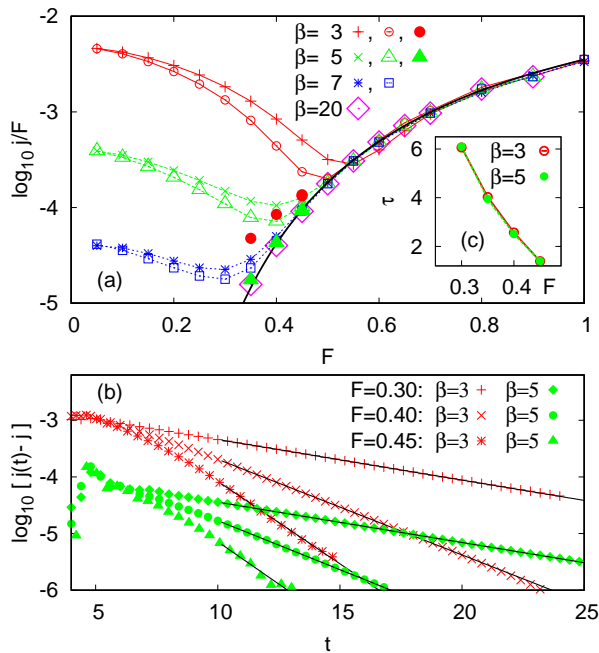}}
 \vspace*{-2mm}
 \caption{
  (a) Time-averaged current, $\jstat/F$, for $U=5$: Cross-like and open symbols correspond 
  to a time average for $8<t<10$ and $10<t<12$, respectively. Filled symbols result 
  from an extrapolation to $t=\infty$ [see panel (b)]. The solid black 
  line is obtained by fitting the data for $\beta=20$ with Eq.~(\ref{fth}) 
  [$F_\text{th}= 1.92$, $\sigma_\text{tun}^\infty=0.066$]. (b) Long-time evolution of 
  the current for $U=5$, analyzed by a nonlinear least square fit $j(t) = 
  \jstat + \delta j\, \exp(-t/\tau)$ (solid lines); the value $\jstat$ is shown 
  by the filled symbols in (a). (c) The relaxation time $\tau$.
  }
\label{fig03}
\label{fig02}
\end{figure}

A plot of the conductance $\jstat/F$ on a logarithmic scale reveals a crossover from the 
temperature-dependent linear response conductivity $\sigma_\text{dc}(T)$ at small $F$ 
to an almost temperature-independent curve at large $F$ (Fig.~\ref{fig02}a). These data 
suggest that the quasistationary current has a nonzero $T=0$ limit, which reflects the 
ground state decay, or dielectric breakdown of the insulator. In analogy to the 
one-dimensional case \cite{Oka2003a,Oka2010a}, we will refer to this limiting value 
as the tunneling current $j_\text{tun}$.  In fact, the low-temperature data in 
Fig.~\ref{fig02}a can be fit with the same exponential law that determines the ground state 
decay rate in the one-dimensional Hubbard model \cite{Oka2003a,Oka2010a}
\begin{equation}
\label{fth}
j_\text{tun}(F) = F\,\sigma_\text{tun}^\infty \exp(-F_\text{th}/F),
\end{equation}
with a threshold $F_\text{th}$ (solid line in Fig.~\ref{fig02}a).

For small fields and $T>0$, the exponentially small $j_\text{tun}(F)$ is dominated
by the linear response current $\sigma_\text{dc}F$. Surprisingly, however, we find that 
thermal and tunneling current do not simply add up in the stationary state, but
$\jstat/F$ can become much smaller than $\sigma_\text{dc}$. The peculiar minimum at the 
crossover between tunneling and linear response regime in Fig.~\ref{fig02}a arises because
the relaxation to the stationary state becomes slower with decreasing field, such
that an average of $j(t)$ in a  fixed time interval still yields $\sigma_\text{dc} F$ for $F\to0$.
A detailed analysis of the long-time behavior at intermediate $F$ reveals that the 
stationary current is approached via an exponential decay $j(t) = \jstat + \delta j\, 
\exp(-t/\tau)$, where $\jstat$ is closer to $j_\text{tun}(F)$ than it is to the linear 
response current (Fig.~\ref{fig03}b). The decay of the thermal current indicates 
that mobile, thermally excited carriers (and those excited during the switch-on) 
are transformed into immobile excitations in the presence of an electric field, 
such that the linear response current is only visible on a timescale $\propto 1/F$. 
Clearly, such a behavior must be specific to a closed system, whereas in an open system 
the coupling to the environment constantly tends to restore the thermal state. Nevertheless, 
the observed relaxation phenomenon is an interesting topic for
future theoretical transport investigations, and it is an open question whether it is a 
generic feature in closed systems.

\begin{figure}
\includegraphics[angle=0,width=0.95\columnwidth]{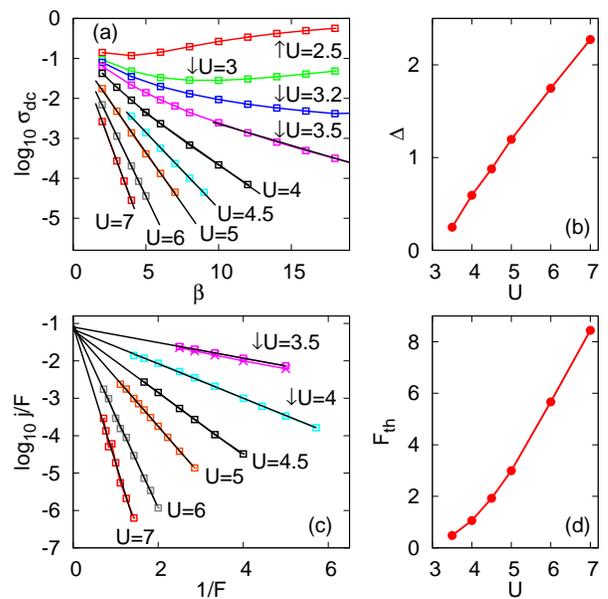}
 \vspace*{-4mm}
 \caption{
  (a) Linear response 
  conductivity, obtained by extrapolating $\jstat/F$ to $F=0$. Thin solid lines
  correspond to fits with Eq.~(\ref{delta}); the resulting gap $\Delta(U)$ 
  is shown in (b). (c) Conductance $\jstat/F$ for the stationary current at 
  $\beta=20$. For $U=3.5$, there is a still a slight drift of the current at 
  the largest times (see text), and we plot time-averages for $8<t<10$ 
  (crosses) and $12<t<14$ (open symbols). Solid lines are linear fits 
  according to Eq.~(\ref{fth}). (d) The threshold field $F_\text{th}(U)$
  resulting from the fits in (c). }
\label{fig04}
\end{figure}

Figure \ref{fig04} summarizes the main numerical results of this paper by comparing the 
current 
in the tunneling regime ($T\downarrow0$) and in the linear response regime ($F\to 0$, 
$t \lesssim 1/F$). For $U \lesssim 3$, the linear response conductivity $\sigma_\text{dc}$ 
increases with decreasing temperature, while it becomes exponentially small for $U\gtrsim3$, 
\begin{equation}
\label{delta}
\sigma_\text{dc} \sim  \exp(-\Delta/T),
\end{equation}
thus signaling the metal-insulator transition at $U\approx 3$ (Fig.~\ref{fig04}a and b).
(Strictly speaking, the metal-insulator transition displayed in Fig.~\ref{fig04} is 
only a crossover because temperatures are larger than $T_c$.) On the other hand, by fitting 
$\jstat/F$ for $\beta=20$ with Eq.~(\ref{fth}), we obtain the threshold $F_\text{th}$ 
as a function of the interaction (Fig.~\ref{fig04}c and d). Deep in the insulating 
phase, both $\Delta$ and $F_\text{th}$ increase linearly with $U$, while they vanish 
as the metal-insulator transition is approached. Because 
the metal-insulator transition is first order in DMFT \cite{Georges1996}, 
observables in the insulating phase are expected to display nonanalytic behavior not at 
the actual transition $U_{c2}$ at $T=0$, but at the spinodal point $U_{c1}$ where the 
insulator at $T=0$ disappears. However, the precise behavior of $\Delta$ and $F_\text{th}$ 
at the transition is difficult to obtain within the current approach: As $F_\text{th}$
decreases, a fit of $\jstat/F$ with Eq.~(\ref{fth}) becomes increasingly difficult
because the stationary state is eventually no longer reached within the numerically 
accessible times for $F<F_\text{th}$ due to the slow saturation of $j(t)$ for 
small $F$ (cf.~Fig.~\ref{fig02}c).


Apart from the crossover region at $U < 4$, however, we do observe the behavior described 
by Eqs.~(\ref{fth}) and (\ref{delta}) over a wide range of temperatures and fields. Although 
our setup involves a closed system in which energy is conserved, we expect 
that these DMFT results resemble the \IV-characteristics of a real Mott-insulating 
material for large and small fields, respectively. While the quasistationary state cannot 
exist forever in our setup, it can indeed be stable in a real solid provided the excess 
energy is passed to the lattice. Since the tunneling current turns out to be essentially 
independent of the excitation of the system (cf.~Fig.~\ref{fig01}), differences between 
the quasistationary current in our setup and the stationary current in experiments 
are expected to be of the order of the linear response current (\ref{delta}), which is negligible 
compared to Eq.~(\ref{fth}) for large enough fields. For small fields or high temperatures, on 
the other hand, the open system should recover the linear response behavior which is lost 
in the closed system at long times. Only in between the linear response and the tunneling 
regime one may thus expect that the current depends essentially on details that are 
not accounted for in the purely electronic description.

We find that the ratio $F_\text{th}/\Delta$ is only weakly dependent on the interaction or 
the bandwidth in the insulator, and it should thus give the correct order of magnitude for the 
breakdown field in real Mott insulators as well. Note that this value, i.e., $F_\text{th} 
\approx 2-3 \,\Delta/ea$, is much larger than the temperature-dependent threshold 
which is obtained in the experiments of Ref.~\cite{tag} in connection with a negative 
differential resistance (for SrCuO$_3$, e.g., $F_\text{th} \approx 10^{-4} \Delta/ea$ at 
$T=190K$). The threshold behavior in these one-dimensional materials must thus be of a 
different origin, and indeed collective excitations were proposed in Ref.~\cite{tag}, as 
the temperature dependence of the experimental threshold is similar to what is expected 
for charge-ordered materials. The larger threshold found in our analysis should be observed 
in paramagnetic Mott insulators when other sources of destabilizing the insulator are not 
present.

In conclusion, we have investigated the dielectric breakdown of a Mott insulator in the 
Hubbard model by computing the current in a strong electric field $F$. Our main result 
is the formation of a quasistationary nonequilibrium state with time-independent current, 
which may be called a field-induced metal. In the limit of small temperature, the stationary 
current  resembles the exponential law [Eq.~(\ref{fth})] for the ground state decay-rate in a 
one-dimensional Hubbard model due to many-body Landau Zener tunneling \cite{Oka2003a,Oka2010a}. 
Its value becomes exponentially small below a threshold field which vanishes at the 
metal-insulator transition. Although the threshold field is generally quite large it 
should be experimentally accessible, e.g., using thin films of insulating material between 
metallic leads.

\section*{Acknowledgements}
We thank H.~Aoki, J.~Freericks, M.~Kollar, J.~Kroha, and N.~Tsuji
for insightful discussions. This work was supported by the Swiss National Science 
Foundation  (Grant PP002-118866). TO was supported by a Grant-in-Aid for Scientific 
Research on Innovative Areas "New Frontier of Materials Science Opened by Molecular 
Degrees of Freedom" of MEXT, Japan.


\begin{thebibliography}{}

   \bibitem{tag}
      Y.\ Taguchi, T.\ Matsumoto, and Y.\ Tokura, Phys. Rev. B {\bf 62},  7015  (2000).

   \bibitem{Oka2003a}
   T.~Oka, R.~Arita, and H.~Aoki,
   Phys.~Rev.~Lett. {\bf 91}, 066406 (2003);
   T.~Oka, and H.~Aoki,
   Phys.~Rev.~Lett. {\bf 95}, 137601 (2005).  

\bibitem{Oka2010a}
    T.~Oka and H.~Aoki, Phys.~Rev.~B {\bf 81}, 033103 (2010).

   \bibitem{Imada1998}
     M.\ Imada, A.\ Fujimori, and Y.\ Tokura, Rev. Mod. Phys. {\bf 70}, 1039 (1998).

   \bibitem{EXPorganic}
    Y.\ Tokura, H.\ Okamoto, T.\ Takao, T.\ Tadaoki, and G.\ Saito, Phys. Rev. B {\bf 38}, 2215  (1988);
    F.\ Sawano, I.\ Terasaki, H.\ Mori, T.\ Mori, M.\ Watanabe, N.\ Ikeda, Y.\ Nogami, Y.\ Noda, Nature  {\bf 437}, 522 (2005).

\bibitem{Greiner2002}
    M.\ Greiner, O.\ Mandel, T.\ Esslinger, and T.\ W.\ H\"ansch,
    Nature {\bf 415}, 39 (2002).



  \bibitem{Georges1996}%
    A.\ Georges, G.\ Kotliar, W.\ Krauth, and M.\ J.\ Rozenberg,
    Rev.\ Mod.\ Phys.\ {\bf 68}, 13 (1996).

  \bibitem{Schmidt2002}%
    P.\ Schmidt and H.\ Monien,
    arXiv:cond-mat/0202046 (unpublished).

  \bibitem{Metzner1989a}%
    W.\ Metzner and D.\ Vollhardt,
    Phys.\ Rev.\ Lett.\ {\bf 62}, 324 (1989).
    
  \bibitem{Freericks2006a}%
    J.~K.\ Freericks, V.~M.\ Turkowski, and V.\ Zlati\'c,
    Phys.\ Rev.\ Lett.\ {\bf 97}, 266408 (2006);
    J.~K. Freericks,
    Phys. Rev. B {\bf 77}, 075109 (2008).

    
  \bibitem{Tsuji2009a}
    N.~Tsuji, T.~Oka, and H.~Aoki,
    Phys. Rev. B {\bf 78}, 235124 (2008);
    Phys. Rev. Lett. {\bf 103}, 047403 (2009).


  \bibitem{Eckstein2008a}
  M.~Eckstein and M.~Kollar,
  \newblock Phys. Rev. Lett. {\bf 100}, 120404 (2008).

 
  \bibitem{Eckstein2009a}
  M.~Eckstein, M.~Kollar, and P.~Werner,
  \newblock Phys. Rev. Lett. {\bf 103}, 056403 (2009);
%
  Phys.~Rev.~B {\bf 81}, 115131 (2010).
  
 
  \bibitem{Eckstein2010c}
  M.~Eckstein and P.~Werner,  arXiv:1005.1872.
 
  \bibitem{Pruschke1993a}
  Th.~Pruschke, D.~L.\ Cox, and M.~Jarrell,
  Phys.\ Rev.\ B {\bf 47}, 3553 (1993).

  \bibitem{Strohmeier2010a}
  N.~Strohmaier, D.~Greif, R.~J\"ordens, L.~Tarruell, H.~Moritz, T.~Esslinger,
  R.~Sensarma, D.~Pekker, E.~Altman, and E.~Demler,
  Phys.~Rev.~Lett. {\bf 104}, 080401 (2010).

  \end{thebibliography}
 \end{document}